\documentclass[twocolumn,english,showpacs,superscriptaddress]{revtex4-1}
\usepackage{geometry}
\usepackage{xcolor}
\usepackage{babel}
\usepackage{units}
\usepackage{amsmath}
\usepackage{amssymb}
\usepackage{stmaryrd}
\usepackage{graphicx}
\usepackage{wasysym}
\usepackage{bbold}
\usepackage[bookmarks=false]{hyperref}
\newcommand{\ket}[1]{\vert{#1}\rangle}
\newcommand{\up}{\mid\uparrow\rangle}

\newcommand{\down}{\mid\downarrow\rangle}
\newcommand{\Yb}{$^{171}$Yb$^+$}
\newcommand{\Ba}{$^{138}$Ba$^+$}

\begin{document}
\title{Fast photon-mediated entanglement of continuously-cooled trapped ions \\ for quantum networking}
\date{\today}

\author{Jameson O'Reilly}
\email{Corresponding author: jameson.oreilly@duke.edu}
\email{Present Address: Department of Physics, University of Oregon, Eugene, OR 97331}
\address{Duke Quantum Center, Departments of Electrical and Computer Engineering and Physics, Duke University, Durham, NC 27708}

\author{George Toh}
\address{Duke Quantum Center, Departments of Electrical and Computer Engineering and Physics, Duke University, Durham, NC 27708}
\author{Isabella Goetting}
\address{Duke Quantum Center, Departments of Electrical and Computer Engineering and Physics, Duke University, Durham, NC 27708}
\author{Sagnik Saha}
\address{Duke Quantum Center, Departments of Electrical and Computer Engineering and Physics, Duke University, Durham, NC 27708}
\author{Mikhail Shalaev}
\address{Duke Quantum Center, Departments of Electrical and Computer Engineering and Physics, Duke University, Durham, NC 27708}
\author{Allison Carter}
\email{Present Address: National Institute of Standards and Technology, Boulder CO  80305}
\address{Joint Quantum Institute, Departments of Physics and Electrical and Computer Engineering, University of Maryland, College Park, MD 20742}
\author{Andrew Risinger}
\email{Present Address: Intel Corp., Hillsboro, OR 97124}
\address{Joint Quantum Institute, Departments of Physics and Electrical and Computer Engineering, University of Maryland, College Park, MD 20742}
\author{Ashish Kalakuntla}
\address{Duke Quantum Center, Departments of Electrical and Computer Engineering and Physics, Duke University, Durham, NC 27708}
\author{Tingguang Li}
\address{Duke Quantum Center, Departments of Electrical and Computer Engineering and Physics, Duke University, Durham, NC 27708}
\author{Ashrit Verma}
\address{Duke Quantum Center, Departments of Electrical and Computer Engineering and Physics, Duke University, Durham, NC 27708}
\author{Christopher Monroe}
\address{Duke Quantum Center, Departments of Electrical and Computer Engineering and Physics, Duke University, Durham, NC 27708}
\address{Joint Quantum Institute, Departments of Physics and Electrical and Computer Engineering, University of Maryland, College Park, MD 20742}

\begin{abstract}
We entangle two co-trapped atomic barium ion qubits by collecting single visible photons from each ion through in-\textit{vacuo} 0.8 NA objectives, interfering them through an integrated fiber-beamsplitter and detecting them in coincidence. This projects the qubits into an entangled Bell state with an observed fidelity lower bound of $F>94\%$. We also introduce an ytterbium ion for sympathetic cooling to remove the need for recooling interruptions and achieve a continuous entanglement rate of $250$ s$^{-1}$.
\end{abstract}

\maketitle

Photonic interconnects between quantum processing nodes may be the only way to achieve large-scale quantum computers, and such an architecture has been proposed for the leading qubit platforms \cite{Duan2010, Bernien2013,Axline2018,Young2022}. Using these connections to distribute remote entanglement between computing modules with high rates and near-unit fidelity should enable universal and fully-connected control over a substantially larger Hilbert space, greatly increasing the collective power of the quantum processors \cite{Gottesman1999,Jiang2007,Monroe2014}. Interconnects between quantum memories, even without multi-qubit universal control, also offer diverse opportunities in quantum sensing \cite{Komar2014,Nichol2022}, communication \cite{Nadlinger2022}, and quantum simulation.

Trapped ions are attractive candidates for both quantum computing and networking due to their natural homogeneity, isolation from their environment, and indefinite idle coherence times \cite{Bruzewicz2019}. These advantages, along with decades of technological development, have led to demonstrations of the highest-fidelity state preparation and measurement (SPAM) \cite{Ransford2021} and coherent operations \cite{Ballance2016,Srinivas2021,Clark2021}, all performed in small systems of just one or two ions. Low errors have also been achieved in medium-sized chains \cite{Cetina2022,Chen2023}, with limits due to weaker trap confinement and resulting motional mode-crowding and crosstalk concerns. Alternatively, smaller ion chains can be shuttled between interaction zones \cite{Kielpinski2002,Pino2021}, but transport already dominates the time budget of current systems with up to 32 qubits \cite{Moses2023}. 

Photonic interconnects can avoid the overhead associated with controlling larger chains and finite shuttling speeds, but they rely on probabilistic excitation and photon emission protocols and photon collection efficiencies that have thus far been limited to a few percent. The current state-of-the-art photon-mediated entangling rate between trapped ion qubits is 182 s$^{-1}$ \cite{Stephenson2020}, 
on par with the mean entanglement rate in shuttling architectures \cite{Moses2023}  but 
much slower than typical local entangling rates of 10-100 kHz \cite{Egan2021,Postler2023}. This demonstration was mainly limited by a success probability of $2.18\times10^{-4}$ in each attempt \cite{Stephenson2020} where the leading inefficiency is the use of 0.6 numerical aperture (NA) objectives that only collect $10\%$ of the photons from each ion. A higher success probability of $2.9\times10^{-4}$ has been achieved by surrounding ions with optical cavities, but the requirement of a much lower attempt rate led to a success rate of just 0.43 s$^{-1}$ \cite{Krutyanskiy2023}. 
In these experiments, the attempt rate is limited by initialization steps, including periodic interruptions to recool the ions, as heating from photon recoil can reduce the collection efficiency and cause state measurement errors \cite{Drmota2023}. 

\begin{figure*}[ht]
\begin{centering}
 \includegraphics[width=170mm]{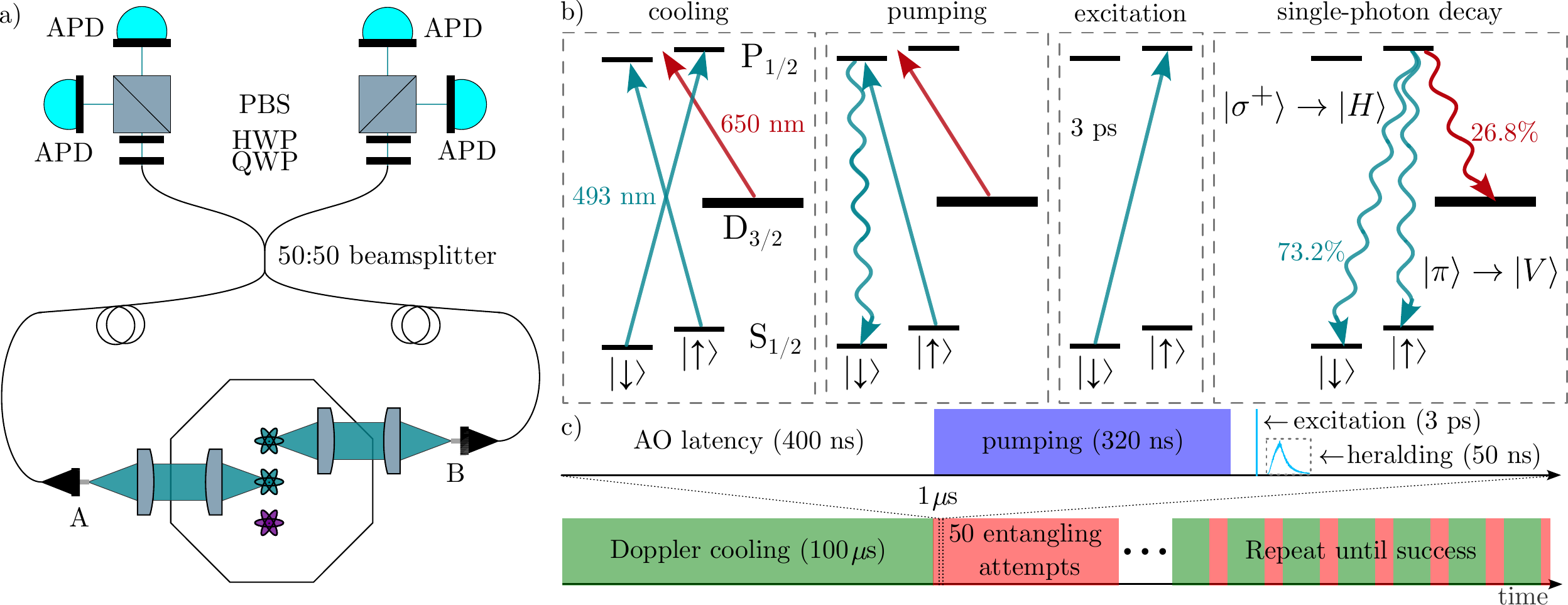}
\par\end{centering}
\centering{}\caption{Overview of the experiment. (a) Three co-trapped ions with two barium ions imaged by in-\textit{vacuo} 0.8 NA objectives \cite{Carter2024} and an optional ytterbium ion for sympathetic cooling. Scattered light at 493 nm is coupled into single mode optical fibers and routed to a Bell state analyzer consisting of an in-fiber beamsplitter to erase which-path information and polarizing beam splitters (PBS) to measure the photon state \cite{Simon2003}. (b) \Ba\ level diagrams for each operation associated with ion-photon entanglement generation. Our qubit states are defined as $\mid\downarrow\rangle\equiv| S_{1/2}, m_J=-1/2\rangle$ and $\mid\uparrow\rangle\equiv| S_{1/2}, m_J=+1/2\rangle$. (c) Timeline for entanglement generation attempts without sympathetic cooling. Each 1 $\mu$s-long attempt consists of optical pumping, pulsed excitation, single photon collection, and fast logic to check for a heralding detection pattern. If no such pattern occurs, we repeat attempts up to 50 times before breaking for Doppler cooling. After cooling, we repeat this cycle until success.}
\label{fig:imagingsystem}
\end{figure*}

In this work, we utilize two 0.8 NA objectives to demonstrate photon-mediated entanglement between \Ba\ ions with a success probability of $2.4(1)\times10^{-4}$ and a fidelity $F\geq93.7(1.3)\%$. Then, we introduce \Yb\ as a sympathetic coolant to achieve an uninterrupted attempt rate of 1 MHz and an ion-ion qubit entanglement rate of $250(8)$ s$^{-1}$. We choose to work with \Ba because it offers the longest-wavelength $S-P$ dipole transition of the commonly-trapped ion species at 493 nm and is similar in mass to \Yb, a well-established species for quantum computing \cite{Cetina2022,Moses2023,Chen2023}. Photons at 493 nm can also be converted to telecom wavelengths for long-distance networking \cite{Hannegan2021}.

We begin by trapping two \Ba\ ions in a four-rod rf Paul trap and Doppler-cooling them with 493 and 650 nm light. Two 0.8 NA in-\textit{vacuo} objectives collect the ion fluorescence with each lens aligned to a different ion and $<10^{-5}$ crosstalk after coupling into single-mode optical fibers (see Figure \ref{fig:imagingsystem}(a)). The trapping and imaging systems are described in more detail in Ref. \cite{Carter2024}. The fiber positions are optimized for maximum coupling before each measurement and are typically stable across multiple hours.

To generate entanglement between each ion and its emitted photon, we begin by optically pumping each \Ba\ ion to $\down\equiv| S_{1/2}, m_J=-1/2\rangle$ and then exciting to $| P_{1/2}, m_J=+1/2\rangle$ with near-unit probability using a 3 ps pulse of $\sigma^+$ 493 nm light produced by a frequency-doubled mode-locked Ti:Sapphire laser \cite{moehring_entanglement_2007} (see Supplemental Material \ref{mira}).
When the ion returns to the $S_{1/2}$ state after spontaneous emission (lifetime $\sim 8$ ns), it can decay to either $\down$ or $\up\equiv| S_{1/2}, m_J=+1/2\rangle$, correlated with the photon polarization. 
When a 493 nm photon is collected perpendicular to the magnetic field axis and coupled into a single-mode fiber, the photon and its parent ion are projected to the state
\begin{equation}
\frac{\ket{H}\down+\ket{V}\up}{\sqrt{2}}, \label{eqn:IP}
\end{equation}
where $\ket{H}$ and $\ket{V}$ represent orthogonal polarizations. The static phase of the above superposition is set to zero for convenience and without loss of generality. 

If a single photon is detected during the 50 ns photon detection window after excitation, we proceed to state analysis and detection. Otherwise, we either repeat the attempt or break for 100 $\mu$s of Doppler cooling after 50 successive attempts, for a duty cycle of $33\%$. Each attempt takes 1 $\mu$s, dominated by AOM latency and state preparation, and has independent single-photon success probabilities of $\eta_A=2.3(1)\%$ and $\eta_B=2.2(1)\%$ (see Supplemental Material \ref{colleff}) through each of the two ion imaging systems (hereafter labelled $A$ and $B$). The full experimental sequence is displayed in Figure \ref{fig:imagingsystem}(c), which is not to scale. 

\begin{figure}[ht]%
\centering
\includegraphics[width=0.47\textwidth]{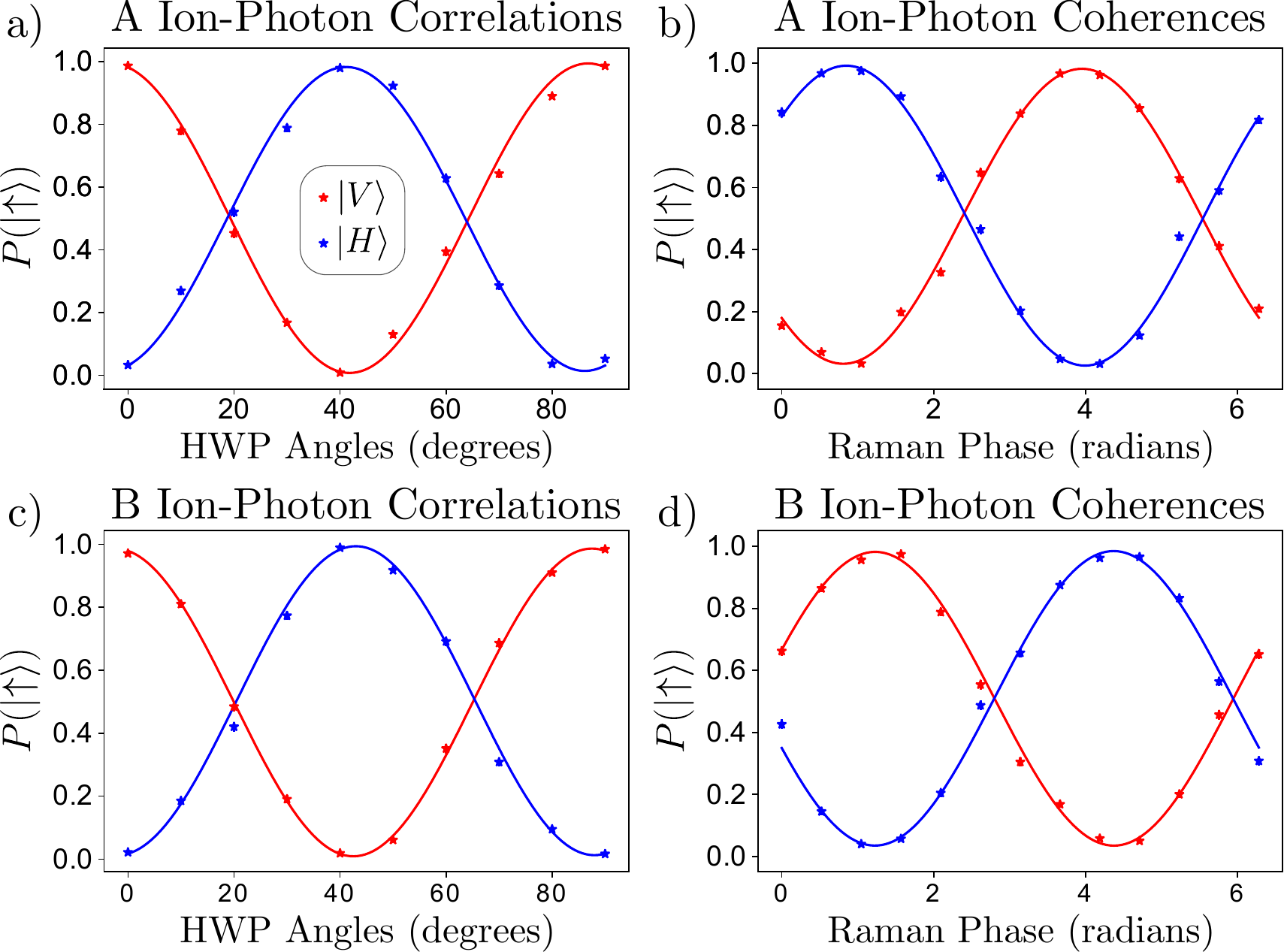}
\caption{Characterization of ion-photon entanglement using imaging systems $A$ and $B$, each coupled to a unique ion. Each data point is based on 200 detection events, making the statistical error bars too small to be visible. All data was taken over the course of about 10 minutes. Red and blue data points correspond to the probability that the ion is in the bright state $\mid\uparrow\rangle$ after a photon is detected in the $V$ or $H$ output mode of a polarizer, respectively. The solid lines are fits to sinusoidal functions.}
\label{fig:corrcoh}
\end{figure}

After the photon exits the fiber, it passes through a quarter-wave plate (QWP) to compensate for any ellipticity. Then, we examine the ion-photon correlations (Figs. \ref{fig:corrcoh}a,c) by scanning the angle of a half-wave plate (HWP) in the beam path and measuring both the photon polarization and the parent ion qubit state (see Supplemental Material \ref{detection}). The resulting contrast in the correlation sets an upper bound on the fidelity overlap with Eq. \ref{eqn:IP} of $F_A<99.1(1)\%$ and $F_B<99.1(7)\%$, which we attribute to residual polarization mixing in the imaging systems. We perform the measurements for each imaging system by physically blocking the other. 

To establish a lower bound for the fidelity of each ion-photon pair, we shrink the photon detection window to 3 ns and rotate each HWP by $22.5^{\circ}$ so that single photon detections herald each parent ion into the equal superposition states
\begin{equation}
\frac{\mid\downarrow\rangle_j\pm e^{i\phi_j}\mid\uparrow\rangle_j}{\sqrt{2}},
\end{equation}
where the sign depends on which detector the photon hits and the phase $\phi_j$ $(j=A, B)$ is given by static polarization rotations in the fiber. Then, we use a pair of 532 nm Raman beams to drive a $\pi/2$ rotation of the atomic qubit with variable phase (Figs. \ref{fig:corrcoh}b,d)  \cite{Crocker2018}. The contrast of the qubit state population with this phase sets a lower bound on the ion-photon fidelities of $F_A>98.1(1.4)\%$ and $F_B>96.8(6)\%$ \cite{Auchter2014}. We also measure unmatched superposition phases of $\phi_A=5.00(2)$ rad and $\phi_B=0.48(2)$ rad caused by different uncompensated birefringence along the two photon paths.

Based on the measured qubit coherence time of $T_2^*=550(27)$ $\mu$s, limited by magnetic field fluctuations, we attribute $0.26(3)\%$ of each infidelity to decoherence during the 40 $\mu$s before the analysis $\pi/2$ pulse. Another $0.10(2)\%$ comes from averaging over different ion qubit phases at the start of the analysis pulse due to the spread of photon detection times within the detection window. We bound errors from double excitations, crosstalk between the imaging systems, and excitation laser background to the $10^{-5}$ level by measuring the ratio between one and two photon events.

Having established ion-photon entanglement through each imaging system, we can now entangle the two ions by sending both photons into a Bell state analyzer as shown in Figure \ref{fig:imagingsystem}, thereby performing entanglement swapping. An in-fiber 50:50 beamsplitter erases the ``which-path" information, so if we detect one $H$ and one $V$ photon in the same trial the ions are ideally heralded into the entangled state
\begin{equation}
\frac{\down_A\up_B \pm e^{i(\delta t+\phi)}\up_A\down_B}{\sqrt{2}} \label{eqn:Bell}.
\end{equation}
Here, the sign is determined by whether a coincident detection occurred on the same or opposite sides of the beamsplitter \cite{Simon2003}, $\delta\equiv\omega_B-\omega_A= 2\pi \times 984(2)$ Hz is the qubit frequency difference between the ions, $t$ is the time elapsed after coincidence detection, and $\phi\equiv\phi_B-\phi_A$. This state suppresses the effect of common-mode noise \cite{Lidar1998,Monz2009} and we indeed measure an extended Bell state coherence time of $T_2^*=38(13)$ ms.

For this experiment, we measure the probability to generate one of the above maximally-entangled states to be $2.4(1)\times10^{-4}$, which is consistent with the product of the measured ion-photon efficiencies above: $\frac{1}{2}\eta_A\eta_B=2.50(16)\times10^{-4}$, with the factor of 1/2 stemming from heralding only two of the four Bell states. The effective attempt rate of 333 kHz is the same as in the individual ion-photon measurements above, so the ion-ion entanglement rate is 
$79(3)$ s$^{-1}$. 

\begin{figure}[t]
\centering
\includegraphics[width=0.37\textwidth]{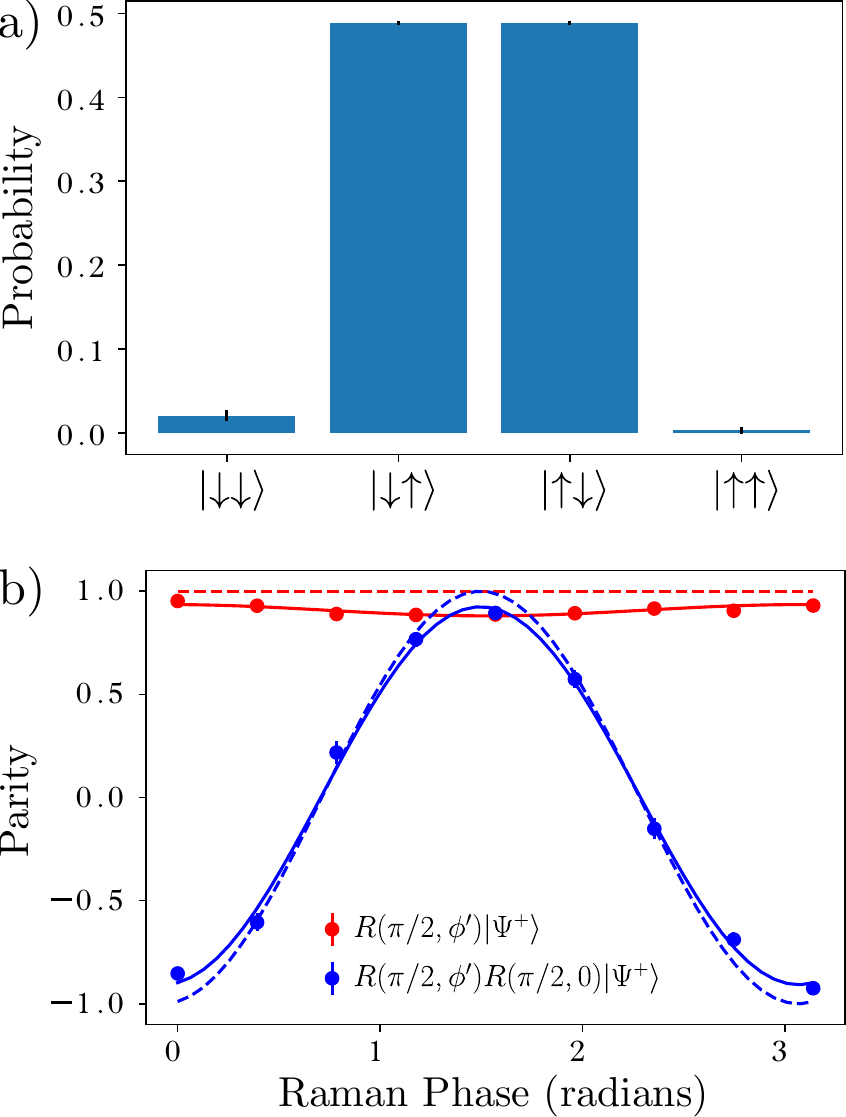}
\caption{Ion-ion entanglement fidelity estimation based on a total of 9,649 entanglement attempts accrued over the course of 40,834,498 attempts and about 8 minutes of clock time. (a) Two-ion state populations. Our detection method cannot distinguish between $\mid\downarrow\uparrow\rangle$ and $\mid\uparrow\downarrow\rangle$ but we assume here that they contribute equally to the measured one-bright population. (b) Parity scan for bounding the off-diagonal elements of the ion-ion state. The red data correspond to scanning the phase of a single $\pi/2$ pulse. The corresponding solid line is the fit and the dashed line represents the expected behavior of the ideal state $\mid\Psi^+\rangle$. The blue data correspond to scanning the phase of a second $\pi/2$ pulse after a $\pi/2$ pulse with fixed phase $\phi=0$.}
\label{fig:ionionanalysis}
\end{figure}

We measure both the populations and coherences of the heralded state of the ions by applying appropriate qubit rotations to both ions, as described in Supplemental Material \ref{bounds}. We measure the populations of the odd parity states to be $P_{\downarrow\uparrow}+P_{\uparrow\downarrow}=97.6(5)\%$ with coherences $2\text{Re}\left(\rho_{\downarrow\uparrow,\uparrow\downarrow}+\rho_{\downarrow\downarrow,\uparrow\uparrow}\right)=92.5(1.7)\%$ \cite{Sackett2000,Slodicka2013}. Bounding the other possible coherence terms results in a SPAM-corrected fidelity with respect to Eq. \ref{eqn:Bell} of $F\geq93.7(1.3)\%$ \cite{Casabone2013}. These results are based on the data displayed in Fig. \ref{fig:ionionanalysis}.

Based on the measured finite contrast of the spin-polarization correlations, we expect an infidelity of $2.9(1.6)\%$, which is consistent with the measured populations. The extended two-qubit coherence is expected to contribute $0.3(1)\%$ and other sources including temporal mismatch and dark counts account for $<0.3(1)\%$. The total predicted infidelity of $<3.5(1.6)\%$ (see Table \ref{table:ionioninfidelity}) is thus within error of our measured infidelity. Notably, using an in-fiber beamsplitter avoids the percent-level error induced by imperfect free-space photon spatial mode overlap in prior experiments \cite{Hucul2015,Stephenson2020}.

\begin{table}
\begin{center}
\begin{tabular}{ |c|c|c| }
 \hline
 Error source & Infidelity ($\times10^{-2}$) \\ 
  \hline
 Polarization errors & $2.9(1.6)$ \\
 Ion decoherence ($T_2^*=38(13)$ ms) & $0.3(1)$ \\
 Timing mismatch ($\Delta t=21(2)$ ps)& $0.13(1)$ \\
 Imperfect splitting ratio ($r=0.515(1)$) & $0.09(1)$ \\  
 Dark counts and double excitations & $<0.1$ \\  
 \hline
 Total & $<3.5(1.6)$ \\
 \hline
\end{tabular}
\caption{Infidelity budget for the entangled ion-ion state. The total expected error is consistent with our measured fidelity lower bound of $F>93.7(1.3)\%$.}
\label{table:ionioninfidelity}
\end{center}
\end{table}


\begin{figure}[ht]
\begin{centering}
 \includegraphics[width=0.47\textwidth]{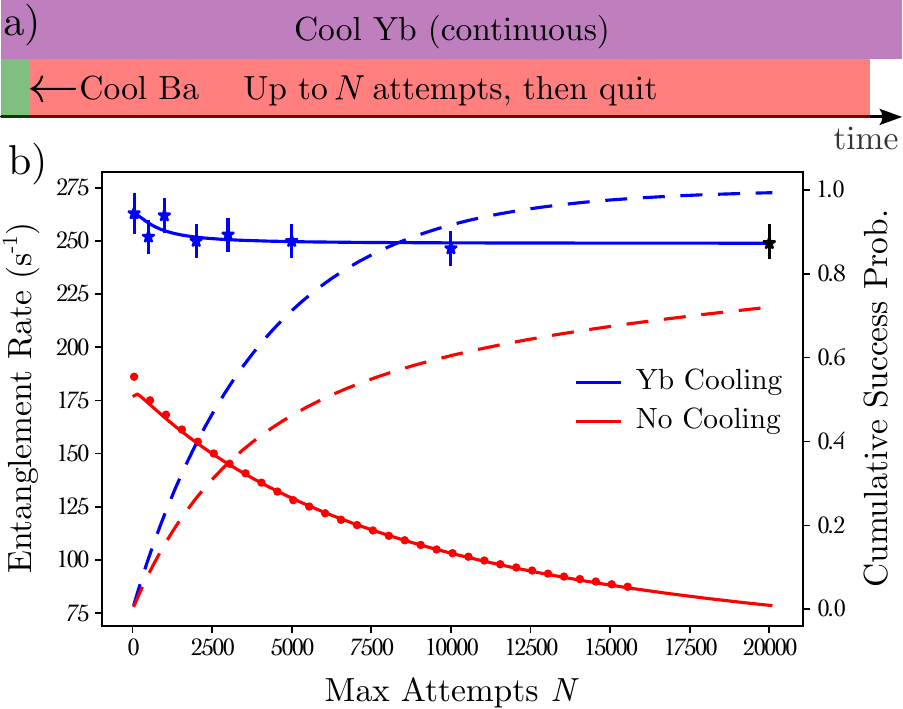}
\par\end{centering}
\centering
\caption{(a) New experimental sequence with an ytterbium sympathetic coolant present in the Coulomb crystal. For each requested event we initialize by Doppler cooling the barium ion for 100 $\mu$s and then execute attempts until we either succeed or reach $N$ failures. Meanwhile, we continuously Doppler cool the ytterbium ion. (b) The points (data) and solid lines (fits) represent average entanglement rates for different maximum consecutive attempts with (blue) and without (red) the ytterbium sympathetic coolant. Red data is calculated from a histogram of attempts before success in 13,739 loops and each blue point is a direct measurement based on an average of 3,741,332 attempts. For a more direct comparison, we ignore barium recooling time in the no cooling case. The final, black point is based on a fit to the blue data. Dashed lines are the cumulative success probability of a heralding event through attempt $N$ for the fitted probability distributions of the Yb-cooling and no-cooling runs.}
\label{fig:ybrescue}
\end{figure}

Over the course of many consecutive entanglement generation attempts, recoil from optical pumping and pulsed excitation heats the ions, reducing the heralded success probability, as shown in Figure \ref{fig:ybrescue}(b). Above, we capped the number of attempts without cooling at 50 to avoid this decay and to maintain high-fidelity state detection. We avoid these issues while maximizing the entanglement attempt rate by co-trapping a \Yb\ ion for continuous sympathetic cooling.

We Doppler cool the \Yb\ ion using 370 and 935 nm light, with sufficient spectral isolation as not to degrade the \Ba\ state detection, cooling, or coherent operations. The relatively similar masses of barium and ytterbium and the small ratio of radial to axial confinement in the trap enable significant coupling between the radial modes of the different species \cite{Sosnova2021}, which in turn allows for efficient sympathetic cooling \cite{sakrejda_efficient_2021}.



With continuous sympathetic cooling, we are able to perform entanglement attempts without stopping for recooling, recovering our full attempt rate of 1 MHz. Although our hardware counter resets at $2^{14}=16384$ attempts, we estimate an average success probability of $2.50(8)\times10^{-4}$ when allowing a maximum of $N=20,000$ attempts, corresponding to an entanglement rate of $250(8)$ s$^{-1}$. After 20,000 attempts, the cumulative success probability of heralding an entanglement event is $>99\%$ (see Supplemental Material \ref{PDF}).




This rate, made possible by increased numerical aperture and the introduction of sympathetic cooling, surpasses any previous mark in a system with Bell state fidelities above $70\%$ \cite{Stockill2017}. 
It could be improved by almost a factor of three by replacing AOMs with electro-optic control to reduce latency and by another factor of two by switching to an atomic species with a larger branching ratio to the ground state \cite{Hucul2015,Stephenson2020}. Building a duplicate of this system and using both imaging systems of each chamber to collect light from a single ion would again double the success probability reported here, providing a road map to kHz-level remote entanglement rates between atomic memories. Further increases could be achieved using Purcell enhancement in short optical cavities or large-scale spatial multiplexing with integrated optics \cite{Ghadimi2017}.

Achieving a better understanding of polarization mapping errors will require a more careful study of imaging aberrations and inhomogeneous birefringence in the lenses and vacuum window. This will be enabled by performing full tomography of the entangled ion-photon and ion-ion states. The dominance of imperfect polarization encoding in our error budget suggests that alternative photonic-qubit encodings, such as frequency \cite{Maunz2009} and time-bin \cite{Ward2022,Tchebotareva2019}, may be beneficial for short and medium-distance networking in addition to their usual application across longer distances \cite{Bersin2024}. The former could be available using the $^{137}$Ba$^+$ or $^{133}$Ba$^+$ isotopes while the latter benefits from the long $D$ state lifetimes in any barium isotope. 

The continued maturation of photonic interconnects will enable a wide variety of quantum technologies including scalable ion-trap quantum computers \cite{Monroe2014}, quantum-limited sensing networks \cite{Komar2014,Nichol2022}, secure communication \cite{Nadlinger2022}, and blind quantum computation \cite{fitzsimons_private_2017}. Many of these applications will require the integration of remote and local entangling operations, both of which benefit from the introduction of sympathetic cooling. Dual-species or \textit{omg} \cite{Allcock} operation is already necessary in most trapped-ion computing and networking architectures and has been demonstrated in numerous experiments \cite{Cetina2022,Drmota2023,Moses2023}, but the integration of these techniques into a multi-node network, with demonstrations of quantum repeaters and entanglement distillation, remains outstanding.



This work is supported by the DOE Quantum Systems Accelerator (DE-FOA-0002253) and the NSF STAQ Program (PHY-1818914). J.O. is supported by the National Science Foundation Graduate Research Fellowship (DGE 2139754).

\bibliographystyle{apsrev4-1}
\bibliography{cleocleo}

\section*{Supplemental Material}

\subsection{Pulsed excitation}
\label{mira}
To produce pulsed 493 nm light, we use a mode-locked Coherent Mira 900P Ti:Sapphire laser at 986 nm and subsequently frequency-double the light using second harmonic generation (SHG) to make 493 nm. The laser generates 3 ps pulses with a repetition rate of 76 MHz. The train of 986 nm pulses enters an electro-optic pulse picker, which transmits single, on-demand pulses with an extinction ratio of about 500:1. After frequency doubling with a MgO-doped, periodically-poled lithium niobate crystal, this extinction ratio increases to >100,000:1. Finally, we send the pulses through an AOM for further extinction and power control before routing them to the vacuum chamber via polarization-maintaining optical fiber.

\subsection{Photon collection efficiencies}
\label{colleff}
In each attempt, we pump to $\mid\downarrow\rangle$ with $96(2)\%$ fidelity and excite an average of $96(2)\%$ of the population to $|P_{1/2}, m_J=+1/2\rangle$. Based on the branching ratio back to $S_{1/2}$, a 493 nm photon is emitted in  $73.2\%$ \cite{Munshi2015} of decay events. The photons are collected by a 0.8 NA objective that covers $20\%$ of the emission solid angle, but within that area only $97(1)\%$ of the photons make it past the trap rods and the lens has a transmission of $91(3)\%$ \cite{Carter2024}. We measure a fiber coupling efficiency of $30(3)\%$ \cite{Carter2024} and detect photons with avalanche photo-detectors that have specified quantum efficiencies of $71\%$. In total, from either imaging system, we expect a single photon detection in $2.5(3)\%$ of trials, which is consistent with our measured values of $2.3(1)$ and $2.2(1)\%$. We believe that the measured values are a bit lower due to ion recoil heating and additional photonic losses from polarizers, waveplates, and optical filters.

\subsection{Two-ion state detection}
\label{detection}
\begin{figure*}[htbp]
\begin{centering}
 \includegraphics[width=170mm]{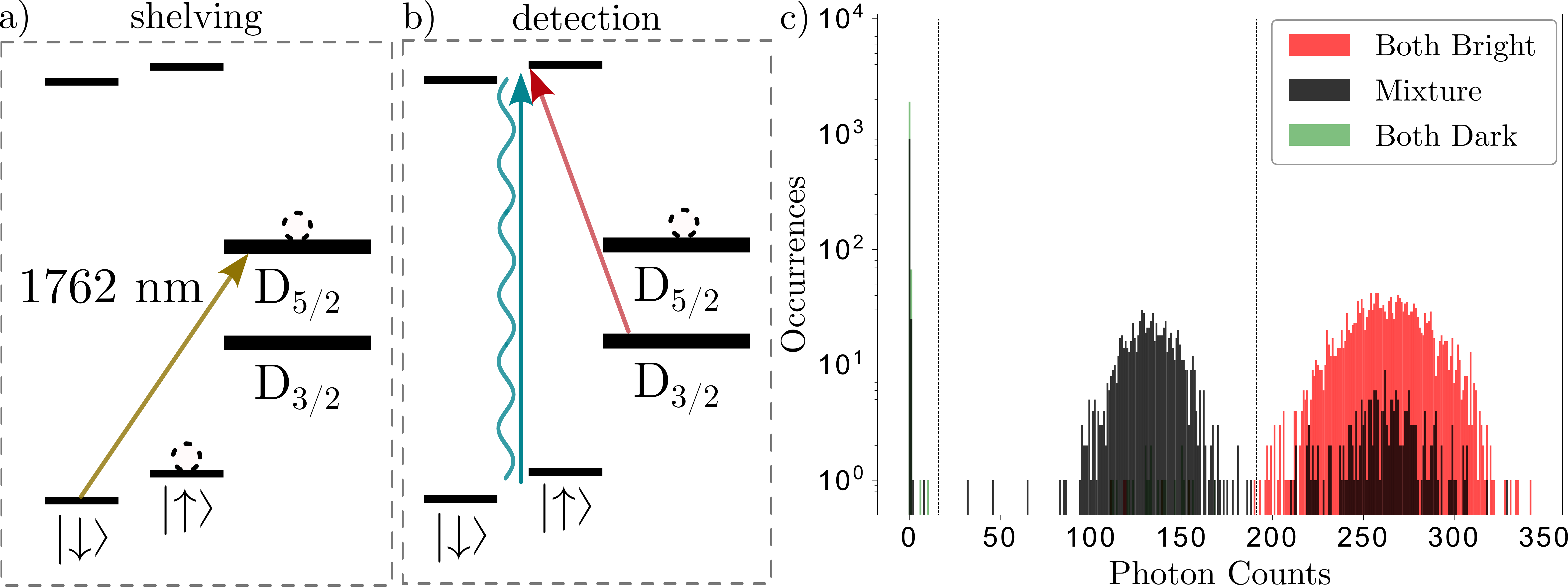}
\par\end{centering}
\centering{}\caption{Two-ion state detection. (a) We shelve the $\mid\downarrow\rangle$ state of each ion to the metastable $D_{5/2}$ manifold. (b) When we subsequently apply 493 and 650 nm light, the shelved population remains dark while any $\mid\uparrow\rangle$ population fluoresces. (c) We collect 493 nm scattered photons from the ions for 1 ms using the imaging system shown in Figure 1 in the main text and set threshold values that separate the photon count distributions for zero, one, or two bright ions with high fidelity. In this example, both bright corresponds to pumping both ions to $\mid\uparrow\rangle$, both dark corresponds to pumping both to $\mid\downarrow\rangle$ and shelving, and mixture is the result of applying a partial shelving pulse so that we get significant one-bright population. We repeated each experiment 20,000 times.}
\label{fig:globaldetection}
\end{figure*}

We detect the state of the qubit(s) at the end of an experiment by shelving the $\mid\downarrow\rangle$ population in the $|D_{5/2},m_J=-1/2\rangle$ state. We shelve using 1762 nm light produced by a thulium-doped fiber laser and fiber amplifier. This system produces 450 mW of 1762 nm light and is stabilized to $<200$ Hz by locking to a high-finesse optical cavity with an ultra-low expansion (ULE) glass spacer. After shelving, we apply all polarizations of 493 and 650 nm light, which causes unshelved ions in $\mid\uparrow\rangle$ to fluoresce while shelved ions remain dark in the metastable $D_{5/2}$ manifold, see Figure \ref{fig:globaldetection}(a-b). We collect and detect these fluorescence photons using the imaging system shown in Figure 1 in the main text \cite{Dietrich2009,Keselman2011}.
For the one-ion case, as in the ion-photon measurements, we achieve an average ion qubit detection fidelity of $\sim99.5\%$.

Collecting fluorescence for 1 ms provides well-resolved photon number histograms for the cases of no bright ions, one bright ion, and two bright ions, as shown in Figure \ref{fig:globaldetection}(c). Imperfect shelving reduces the no-bright detection fidelity to $98.7(4)\%$ with erroneous events predominantly registering as one-bright events. The decay of fiber coupling due to heating during experiments reduces the two-bright average fidelity to $98.1(4)\%$. We correct for these and their corresponding single-ion errors by applying the inverse transformations to the data \cite{Shen2012,Seif2018}.

\subsection{Ion-ion entanglement fidelity bound}
\label{bounds}
The global nature of our Raman addressing system limits us to analyzing the state fidelity relative to the state
\begin{equation}
\ket{\Psi^+}\equiv \frac{\down\up + \up\down}{\sqrt{2}}
\end{equation}
because the singlet state $|\Psi^-\rangle$ is invariant under global rotations. We begin this process by measuring the state populations $\rho_{\downarrow\uparrow}+\rho_{\uparrow\downarrow}=97.6(5)\%$. 
After waiting 210 $\mu$s for $\delta t=-\phi$, we apply a global $\frac{\pi}{2}$ rotation that converts $|\Psi^+\rangle$ into $|\Phi^+\rangle=\frac{1}{\sqrt{2}}\left(\mid\downarrow\downarrow\rangle+\mid\uparrow\uparrow\rangle\right)$ followed by a second $\frac{\pi}{2}$ pulse with varying phase $\phi'$. The maximum of the parity $P\equiv\tilde{\rho}_{\downarrow\downarrow}+\tilde{\rho}_{\uparrow\uparrow}-\tilde{\rho}_{\downarrow\uparrow}-\tilde{\rho}_{\uparrow\downarrow}$ of the rotated state in this scan, shown in Figure 3 in the main text, corresponds to $2\text{Re}\left(\rho_{\downarrow\uparrow,\uparrow\downarrow}+\rho_{\downarrow\downarrow,\uparrow\uparrow}\right)=92.5(1.7)\%$ \cite{Sackett2000}. We measure $2\text{Re}(\rho_{\downarrow\downarrow,\uparrow\uparrow})=2.7(1.8)\%$ by scanning the phase of a single $\frac{\pi}{2}$ pulse \cite{Slodicka2013}, which allows us to calculate an ion-ion fidelity lower bound of $F\geq93.7(1.3)\%$ \cite{Casabone2013}.

\subsection{Yb-Ba-Ba collective motional modes}
\label{modes}
Coulomb forces between ions co-trapped in a harmonic potential $U$ lead to collective motional modes that are often used as an information bus for local entangling gates in trapped ion systems \cite{James1998,Sorensen2000,SchmidtKaler2003}. In our application, the collective nature allows us to cool the barium ions via their coupling to the ytterbium ion's motion. Using our measured secular frequencies for a single barium ion \cite{Carter2024}, we can find the structure of the modes in our Yb-Ba-Ba chain by solving
\begin{equation}
    \sum_{i,j=1}^N\frac{\partial U}{\partial q_i\partial q_j}\bigg|_0b_{im}=\omega_m^2m_ib_{im}
\end{equation}
where $q_i$ is the position of ion $i$, $\omega_m$ is the secular frequency of mode $m$, and $b_{im}$ is the participation eigenvector of ion $i$ in mode $m$ with $\sum_i b_{im}b_{in} = \delta_{nm}$ and $\sum_m b_{im}b_{jm} = \delta_{ij}$ .


\begin{table}
\begin{tabular}{ |c|c c c|c| }
 \hline
 & \Yb & \Ba & \Ba & $\omega_m/2\pi$ (kHz) \\ 
  \hline
 Axial Mode 1 & 0.614 & 0.640 & 0.300 & 353 \\
 Axial Mode 2 & 0.567 & -0.126 & -0.840 & 604 \\
 Axial Mode 3 & 0.549 & -0.758 & 0.453 & 872 \\
 Radial Mode 1 & 0.178 & 0.412 & 0.847 & 868 \\
 Radial Mode 2 & 0.587 & 0.672 & -0.512 & 737 \\
 Radial Mode 3 & 0.790 & -0.615 & 0.144 & 606 \\
 \hline
\end{tabular}
\caption{Collective secular motional mode participation eigenvector matrix $b_{im}$ of the Yb-Ba-Ba Coulomb crystal that we trap for sympathetic cooling experiments. Ions with different charge-to-mass ratios typically have good mutual participation in axial modes, and we also maintain strong coupling in the radial modes thanks to our relatively weak radial confinement \cite{Sosnova2021}.}
\label{table:modes}
\end{table}

The excitation and pumping beams are delivered at $45^\circ$ relative to the trap axis and emission is isotropic, so we need to sympathetically cool both the radial and axial directions. Multi-species ion traps often suffer from weak radial coupling between the species \cite{Home2013,Sosnova2021}, but this is circumvented by using a high ratio of axial to radial confinement (see Table \ref{table:modes}).

\subsection{Deriving $\bar{p}(N)$}
\label{PDF}
In a system where the probability of success on the $n^\text{th}$ trial $p(n)\equiv p$ is constant \cite{Drmota2023}, we expect an exponential distribution of required trials $n$ before success each time we attempt to generate entanglement: PDF$(n)=pe^{-np}$. Instead, we observe a success probability that decays to a steady-state value $p(n)=Ae^{-Bn}+C$ (Fig. 4(b) in the main text), stemming from the increased Doppler temperature of the high-intensity optical pumping beam. In this case, we extend the trial number $n$ as a continuous variable and use
\begin{equation}
\text{PDF}(n)=p(n)\left(1-\int_0^n\text{PDF}(n')\text{d}n'\right)
\end{equation}
to find 
\begin{equation}
\text{PDF}(n)=\exp\left[\frac{A}{B}(e^{-Bn}-1)-Cn\right](Ae^{-Bn}+C). 
\end{equation}
Integrating this from 0 to $N$, we find the cumulative density function, or the probability of success up through $N$ trials, 
\begin{equation}
\text{CDF}(N)=1-\exp\left[\frac{A}{B}(e^{-BN}-1)-CN\right].    
\end{equation}
 Finally, we arrive at the average success probability by dividing the probability of success up through $N$ trials by the total attempts that have been executed up to the $N^\text{th}$ in each loop, resulting in
\begin{equation}
\bar{p}(N)=\frac{\text{CDF}(N)}{N+1-\int_0^N\text{CDF}(n)\text{d}n}.
\end{equation}
While we could not find an analytic solution for the integral, we were able to fit the blue data points in Figure 4(c) in the main text to this equation by integrating numerically.


\end{document}